\documentclass[prb,twocolumn,showpacs,superscriptaddress,apsrev4-1]{revtex4-1}
\usepackage[dvips]{graphicx}
\usepackage{gensymb}
\usepackage{amsmath}
\usepackage{amssymb}
\usepackage{mathrsfs}
\usepackage{mhchem}
\usepackage{ulem}
\usepackage[mathscr]{euscript}

\begin{document}

\title{Strain-induced Dirac cone-like electronic structures and semiconductor-semimetal transition in graphdiyne}
\author{Hui-Juan Cui}
\affiliation{School of Physics, University of Chinese
Academy of Sciences, Beijing 100049, China}
\author{Xian-Lei Sheng}
\affiliation{Beijing National Laboratory for Condensed Matter Physics,
 Institute of Physics, Chinese Academy of Sciences, Beijing 100190, China}
\author{Qing-Bo Yan}
\affiliation{College of Materials Science and Opto-Electronic
Technology, University of Chinese Academy of Sciences, Beijing
100049, China}
\author{Qing-Rong Zheng}
\email[]{Email: qrzheng@ucas.ac.cn}
\affiliation{School of Physics, University of Chinese
Academy of Sciences, Beijing 100049, China}
\author{Gang Su}
\email[]{Email: gsu@ucas.ac.cn}
\affiliation{School of Physics, University of Chinese
Academy of Sciences, Beijing 100049, China}

\begin{abstract}
By means of the first-principles calculations combined with the
tight-binding approximation, the strain-induced semiconductor-semimetal transition
in graphdiyne is discovered. It is shown that the band gap of graphdiyne
increases from 0.47 eV to 1.39 eV with increasing the biaxial tensile strain, while the band gap decreases from 0.47 eV to nearly
zero with increasing the uniaxial tensile strain, and Dirac
cone-like electronic structures are observed. The uniaxial strain-induced changes of the electronic structures of graphdiyne come from the breaking of geometrical symmetry that lifts the degeneracy of energy bands. The properties of
graphdiyne under strains are disclosed different remarkably from that of
graphene.
\end{abstract}

\maketitle

Carbon, one of the most common chemical elements in Nature, has a lot of allotropes, plenteous properties and numerous potential
applications. Besides the well-known graphite,
graphene\cite{geim2004}, and diamond, many other novel carbon allotropes, such as T-carbon \cite{Tcarbon}, M-carbon%
\cite{Mcarbon}, graphyne \cite{graphyne,direc},
graphdiyne \cite{graphdiyne}, octgraphene \cite{otc}, etc., are proposed and studied. Owing to its unique electronic \cite%
{elec}, mechanical and thermal properties, graphene stimulated
considerable interest in itself and other two-dimensional (2D) materials
\cite{tubular1,tubular2,tubular3,tubular4}. Graphene's brothers,
graphyne and graphdiyne, are also intriguing
2D carbon materials. In particular, both building blocks and cut-outs were already obtained experimentally \cite{graphdiyne,synthesized,syn2,syn3,syn4,syn5}. It is known that graphene is a zero-gap semimetal, while graphdiyne is a semiconductor with a direct band gap. In addition, graphdiyne
has the high-temperature stability and shows mechanical
properties similar to graphene \cite{carrier}, which is
considered to be a potential material for nanoelectronics.  The
electronic modulation of graphdiyne\cite{gribbons,Mbandgap,MD}is thus of great interest.

\begin{figure}[tbp]
\includegraphics[width=8.0cm]{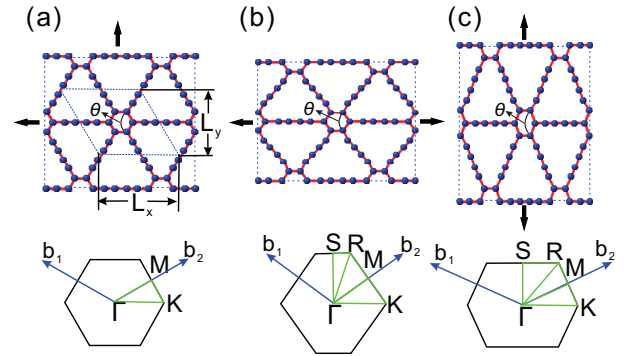}
\caption{(Color online) The schematic structures of graphdiyne under various tensile strains. (a) under biaxial tensile strains, where the parallelogram (dot line)
indicates a unit cell with length $L_{x}$ and width $L_{y}$; (b) under a uniaxial strain along the armchair direction; (c) under a uniaxial strain along the zigzag
direction. The corresponding Brillouin zones are illustrated below each
structure.}
\end{figure}

As graphene has no energy gap, which poses difficulties for wide applications in microelectronic devices, plenty of researches were devoted to
open a band gap in graphene. One method, among others, is to invoke strains
\cite{strain,ribbon,ribbonnew,optical,strainarm,lr,strain2}, which is also
found effective in modulating electronic properties of graphdiyne and other
2D materials such as BN sheet \cite{BN,BN2}. In this work, a
systematic first-principles density-functional theory (DFT) study on
the strain-induced changes of electronic structures in graphdiyne
was performed. In the absence of a strain graphdiyne is known as a semiconductor with a direct band gap. In the presence of strains, we found that the band gap of graphdiyne increases with increasing the biaxial tensile strains, but the band gap decreases with increasing a uniaxial tensile strain either along the armchair or zigzag direction. The electronic properties around the
Fermi surface are dominated by $2p_{z}$ orbitals, and the
Dirac cone-like electronic structures can be obtained. The
calculated electron density of graphdiyne under
different strains shows that C$\mbox{--}$C bonds are elongated, on which the
electron densities are reduced significantly, while
$\ce{{C}\tbond{C}}$ bonds are changed little. We also proposed a tight-binding
model of $\pi$ electrons to describe the low-energy physics around the
Fermi level, which gives the results in well agreement with the DFT calculations. The decline of the band gap of graphdiyne under an asymmetric, uniaxial strain may be owing to the breaking of the geometrical symmetry that lifts the degeneracy of energy bands. These observations in graphdiyne are different from those
of graphene, and may be useful for nanoelectronics.

All calculations were performed by DFT \cite%
{dft1,dft2} implemented within the Vienna \textit{ab initio}
simulation package (VASP) \cite{vasp1,vasp2} with the projector
augmented-wave (PAW) method \cite{paw}, and the results are
rechecked using the Quantum-ESPRESSO package \cite{espresso}. The
generalized gradient approximation (GGA) developed by Perdew and
Wang \cite{ggapw} are adopted for the exchange correlation
potential. The Monkhorst-Pack scheme is used to sample the Brillouin
zone \cite{MPscheme}, and a mesh of $17\times 17\times 1$ $k$-point
sampling was used for the calculations. The total energy was
converged to within $1$ $meV$ with the plane-wave cutoff energy 500
 eV. The geometries were optimized when the remanent
Hellmann-Feynman forces on the ions are less than $0.01$ $eV/nm$. The
distances between two layers in the supercell are more than $10$
${\AA}$\ to avoid the interlayer interactions.

\begin{figure}[tbp]
\includegraphics[width=8.0cm]{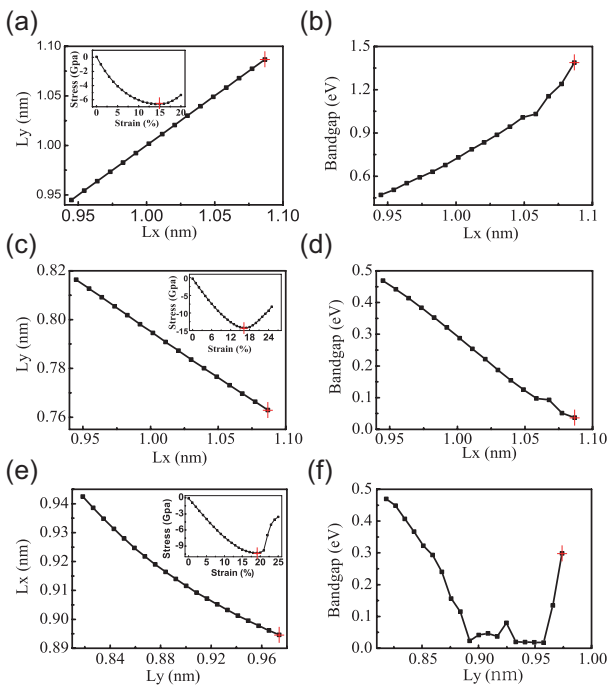}
\caption{(Color online) (a) $L_{y}$
versus $L_{x}$ and (b) the band gap versus $L_{x}$ of graphdiyne under
biaxial tensile strains; (c) $L_{y}$ versus $L_{x}$ and (d) the band gap versus $L_{x}$ of graphdiyne under a uniaxial strain along the armchair direction; (e) $L_{x}$ versus $L_{y}$ and (f) the band gap versus $L_{y}$ of graphdiyne under a uniaxial strain along the zigzag direction. The strain-stress relation for each case is shown in the inset of (a), (c) and (e), respectively. The crosses (red) indicate the largest strain we considered for each case, which can be taken as the transition point from elastic to nonelastic situation.}
\end{figure}

The schematic structure of graphdiyne is shown in Fig. 1(a), which
can be considered as the structure that one third of
C$\mbox{--}$C bonds in graphene are displaced with
two \ce{{C}\tbond{C}} units. These two
\ce{{C}\tbond{C}} units form acetylenic
linkages. The parallelogram in Fig. 1(a) indicates the unit cell of
graphdiyne with length $L_{x}$ and width $L_{y}$. The geometrical and
electronic properties of graphdiyne in the absence of strains were calculated, and
the results reveal that the lattice constants are $L_{x}^{0}=0.945$ $nm$ and
$L_{y}^{0}=0.82$ $nm$, and the a direct band gap is 0.47 eV at $\Gamma $ point,
as shown in Fig. 3(a), which are in good agreement with the previous
calculations \cite{lengths,bandgap}.

Now let us consider graphdiyne under three types of strains: (i) under symmetrical biaxial tensile strains, as indicated in Fig. 1(a); (ii) under a uniaxial tensile strain along the armchair direction, as indicated in Fig. 1(b); and (iii) under a uniaxial tensile strain along the zigzag direction, as indicated in Fig. 1(c). For the case (i), when graphdiyne is stretched under biaxial strains, $L_{x}$ and $L_{y}$ are stretched uniformly in the same ratio, and the strain can be represented as a ratio defined by $(L_{x}-L_{x}^{0})$/$L_{x}^{0}$ in percent. For the cases with uniaxial strains, only one direction is stretched, while another direction is set to free, which makes the whole system reach the lowest energy. Thus, for case (ii) the strain can be measured by $(L_{x}-L_{x}^{0})$/$L_{x}^{0}$ in percent, and for case (iii), the strain can be measured by $(L_{y}-L_{y}^{0})$/$L_{y}^{0}$ in percent. In our calculations, the strain is changed in every $1\%$.

Fig. 2(a) shows $L_{y}$ versus $L_{x}$ and stress-strain relation of
graphdiyne under symmetrical biaxial strains. $L_{y}$ {\it vs.} $L_{x}$ curve gives a raising straight line, and the stress-strain
curve shows a minimum at the strain of $15\%$, which can be viewed as the
transition point from elastic to nonelastic situation for graphdiyne under
the biaxial strains. In this case, the band gap increases from
0.47 eV to 1.39 eV with increasing $L_{x}$, as seen in Fig. 2(b). When the strain is executed along the armchair direction (case (ii)), $L_{y}$ {\it vs.} $L_{x}$ shows a declining straight line, as revealed in Fig. 2(c), which is
similar to that of graphene under a strain perpendicular to C$\mbox{--}$C
bonds (Ref. \onlinecite{strain}), indicating that graphdiyne is the 2D material with a positive Poisson's ratio. The elastic-nonelastic transition occurs
at the strain of $16\%$, close to the value of graphdiyne under biaxial
strains. In this case, the band gap of graphdiyne is decreased almost linearly from 0.47 to 0.02 eV with increasing $L_{x}$ (Fig. 2(d)), which is distinct from the case under biaxial strains. For case (iii), $L_{y}$ as a function of $L_{x}$ shows a concave descending curve, as shown in Fig. 2(e), which is also similar to that of graphene under a strain parallel to C$\mbox{--}$C bonds (Ref. \onlinecite{strain}). The elastic-nonelastic transition occurs around the strain of $20\%$. Fig. 2(f) presents the band gap of graphdiyne under the strain along the zigzag direction, which decreases from 0.47 to 0.02 eV with increasing $L_{y}$ from $0.82$ $nm$ to $0.88$ $nm$. When $L_{y}$ is between $0.90$ $nm$ to $0.95$ $nm$, the band gap is nearly zero, indicating a semiconductor-semimetal transition. When $L_{y}$ is larger than $0.95$ $nm$, the band gap re-opens again. This latter property is also supported by the TB calculations (see below). These properties
are dramatically different from those of graphene \cite{strain,exp,bilayer}.

\begin{figure}[tbp]
\includegraphics[width=8.0cm]{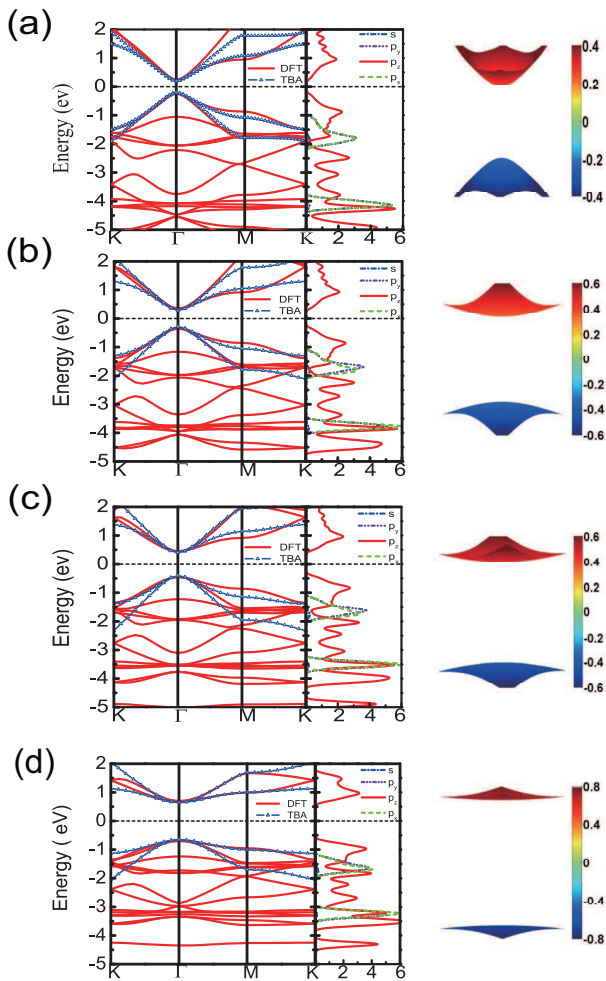}
\caption{(Color online) The band structures calculated by the density functional theory (DFT) and the tight-binding approximation (TBA), the partial density of states (PDOS) and the three-dimensional energy bands at $\Gamma$ point for graphdiyne (a) without a strain, and under a symmetrical biaxial tensile strain of (b) $5\%$, (c) $9\%$, and (d) $15\%$, respectively.}
\end{figure}

\begin{figure}[tbp]
\includegraphics[width=8.0cm]{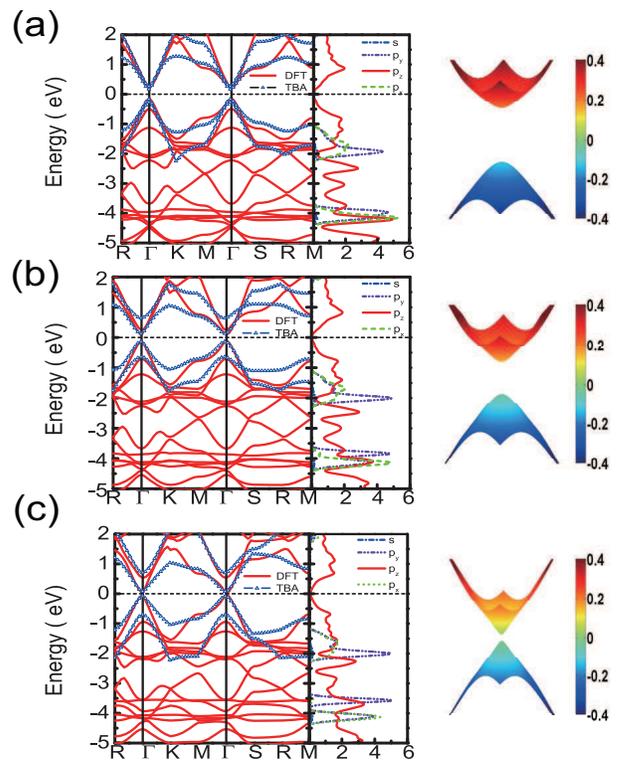}
\caption{(Color online) The band structures calculated by the density functional theory (DFT) and the tight-binding approximation (TBA), the partial density of states (PDOS) and the three-dimensional energy bands at $\Gamma$ point for graphdiyne (a) without a strain, and under a uniaxial tensile strain along the armchair direction with the strain of (b) $5\%$, (c) $9\%$ and (d) $15\%$, respectively.}
\end{figure}

In order to better understand the strain effect on electronic properties
of graphdiyne, the energy bands, partial density of states (PDOS), and electron density distribution under biaxial or uniaxial strains are calculated. Fig. 3(a) shows the electronic structure of graphdiyne in the absence of strain. It is seen that graphdiyne is a semiconductor with a direct band gap 0.47 eV at $\Gamma $ point, which is consistent with previous results \cite{lengths,bandgap}. The degeneracy of energy bands can be observed in both the lowest conduction bands and the
highest valence bands near $\Gamma $ point. The PDOS calculations show that $2p_{z}$
orbitals dominate the states around the Fermi level, while $2p_{x}$ and
$2p_{y}$ orbitals contribute the same, which are mainly located at least 1
eV below the Fermi level. The three-dimensional (3D) energy bands near $\Gamma $ point around the Fermi level reveal two opposite bowls, showing the parabolic-like dispersions. Figs. 3(b)-(d) show the electronic structure of graphdiyne under a
symmetrical biaxial tensile strain of $5\%$, $9\%$ and $15\%$, respectively. It can be observed that for all these cases, the minimal energy gap is still located at $\Gamma$ point, but it gradually becomes larger and larger as
the strain increases. The degeneracy of the lowest conduction bands
and the highest valence bands still maintains, as the geometrical symmetry is
unchanged under the symmetrical strain. Despite of the variation of band gaps, the PDOS remains almost unchanged, and $2p_{z}$ orbitals are still dominating around the Fermi level. From these figures, one may note that the energy bands become more flat as the strain increases, revealing that $\pi$ electrons become more localized
under the symmetrical biaxial strains. This observation is also confirmed in the electron densities shown in Figs. 6(a)-(c), where the electron densities between carbon atoms decrease significantly.

Figures 4(a)-(c) present the electronic structures of graphdiyne
under uniaxial tensile strains along the armchair direction with the strain of $5\%$, $9\%$ and $15\%$, respectively. It is found that the minimal energy gaps are still located at $\Gamma $ point. However, with the increase of strain, the degeneracies of the highest valence bands and the lowest conduction bands are lifted. The energy
levels of the highest valence bands move up and those of the lowest conduction
bands move down, resulting in that the energy gaps diminish with the
increase of strain. Especially, the energy gap becomes very small
(0.02 eV) when the strain is larger than $15\%$. A careful observation
finds that the highest valence bands and the lowest conduction
bands from $\Gamma $ point to $M$ point are separated much
more than the bands from $\Gamma $ point to $S$ point,
and the gap at $M$ point decreases gradually, which is resulted from the
broken symmetry in the Brillouin zone. It is also uncovered in PDOS calculations
that $2p_{z}$ orbitals still dominate the states around the Fermi
level, and the contribution of $2p_{x}$ and $2p_{y}$ orbitals
becomes different. In addition, it is interesting to observe that a Dirac
cone-like electronic structure is formed (Fig. 4(c)).

\begin{figure}[tbp]
\includegraphics[width=8.0cm]{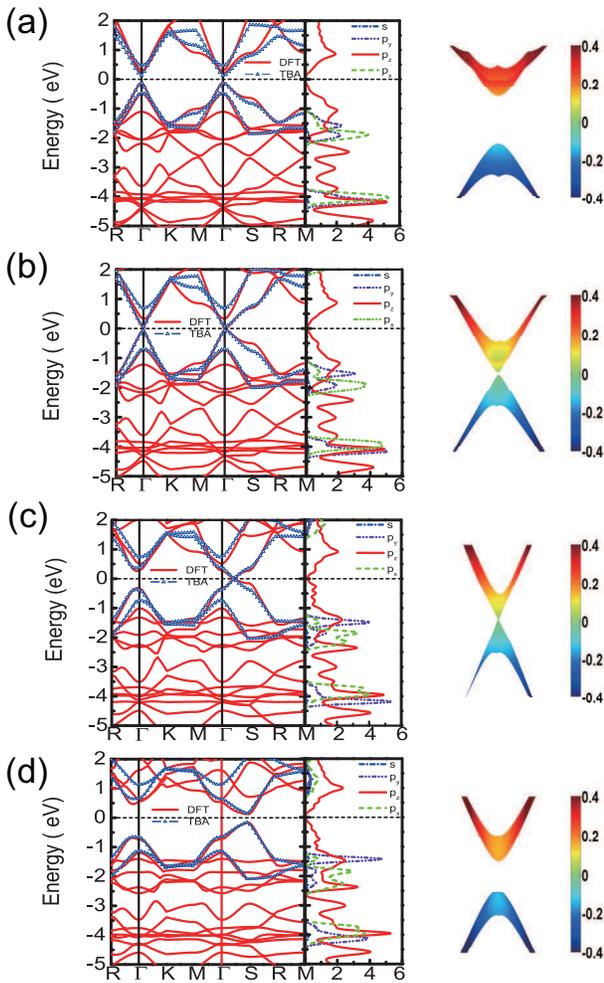}
\caption{(Color online) The band structures calculated by the density functional theory (DFT) and the tight-binding approximation (TBA), the partial density of states (PDOS) and the three-dimensional energy bands for graphdiyne under a uniaxial tensile strain along the zigzag direction with the strain of (a) $5\%$ and (b) $9\%$ at $\Gamma$ point; and (c) $15\%$ and (d) $19\%$ at the $k$ point with minimal energy gap.}
\end{figure}

The electronic structures of graphdiyne under
uniaxial tensile strains along the zigzag direction are more complex, as displayed in
Figs. 5(a)-(d), in which the cases with strains of $5\%$, $9\%$, $15\%$ and $19\%$ are presented. When the strain is increased from zero to
$9\%$ (Figs. 5(a) and 5(b)), the degeneracy of the highest valence bands is
removed, and the minimal band gap opens at $\Gamma$ point and tends to decrease to nearly zero (Fig. 5(b)). When the uniaxial strains between $9\%$ and $17\%$ are applied,
the band gap still maintains nearly zero but the position at which the gap closes is moving gradually from $\Gamma$ point to $S$ point. The corresponding 3D band structure around the Fermi level manifests a Dirac cone-like structure, in which the cone
becomes sharp with the increase of strains, as seen in Fig. 5(c). This observation shows that a strain-induced semiconductor-semimetal transition takes place in graphdiyne. When the strain is larger than $17\%$, the band gap opens again at $S$ point (Fig. 5(d)). It is interesting to note that, in contrast to the case for the strain along the armchair direction, in this case the bands near the Fermi level from $\Gamma $ to $S$ point are lifted much more than the bands from $\Gamma $ to $M$ point, which is due to the broken symmetry of the Brillouin zone.
The PDOS calculations reveal that $2p_{z}$ orbitals dominate the states around Fermi
level, while $2p_{x}$ and $2p_{y}$ orbitals contribute in different ways, which is also owing to the symmetry breaking caused by the asymmetrical uniaxial tensile strains.

\begin{table*}[tbp]
\caption{The lattice constant, plane density $\rho$, energy gap $E_{g}$ between the bottom of conduction band and the top of valence band, Poission's ratio $\upsilon$,
 and the angle $\theta$ shown in Fig. 1(a) under different strains for graphdiyne.}%
\begin{tabular*}{16cm}{@{\extracolsep{\fill}}lcccccccccc}
\hline\hline
strain{$\%$} & lattice constant ({\AA }) & $\rho$ $(mg/m^{2})$ & $E_{g}$ $(eV)$ &  $\upsilon$ & $\theta$ $(^{\circ})$ & $ E_{c}$ $(eV/atom)$
\\ \hline
unstrain (present work) & 9.45 &0.466&0.47& 0.38, 0.476& 120.00& 8.48 \\
unstrain (Ref. \onlinecite{bandgap,otc})& 9.44 &0.466&0.53&0.41, 0.44& 120.00& 7.78 \\
biaxial ($5\%$) & 9.823 &0.426&0.677&$\thicksim$& 120.00 & 8.34\\
biaxial ($9\%$) & 10.30 &0.396&0.888&$\thicksim$& 120.00 & 8.10 \\
biaxial ($15\%$)& 10.87 &0.352&1.389&$\thicksim$& 120.00 &7.58 \\
armchair direction ($5\%$) & 9.397 &0.458&0.321 &0.386& 111.616 &8.45\\
armchair direction ($9\%$) & 9.374 &0.449&0.156& 0.384& 108.213 &8.37 \\
armchair direction ($15\%$)& 9.366 &0.437&0.037&0.377& 107.255 &8.20 \\
zigzag direction ($5\%$) & 9.755 &0.456&0.294&0.434& 127.397 &8.46 \\
zigzag direction ($9\%$) & 10.022&0.445&0.024&0.387 & 131.822&8.36 \\
zigzag direction ($15\%$)& 10.435&0.427&0.019&0.335 & 136.836&8.15  \\
zigzag direction ($19\%$)& 10.717&0.421&0.299&0.308 & 140.198&7.99  \\ \hline\hline
\end{tabular*}%
\end{table*}

\begin{figure}[tbp]
\includegraphics[width=8.0cm]{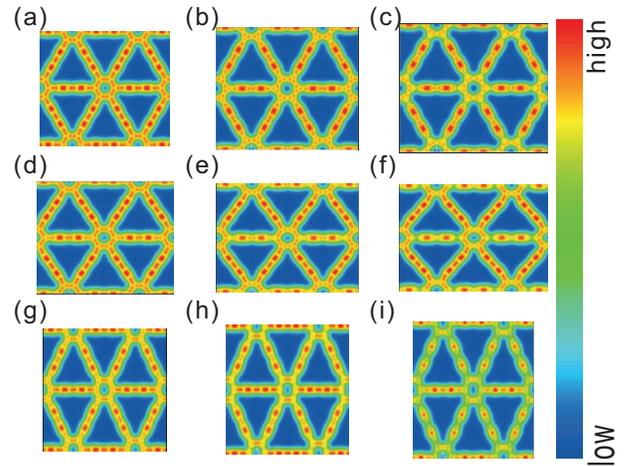}
\caption{(Color online) The distributions of electron densities of
graphdiyne under (a) no strain; biaxial strains of (b) $9\%$ and (c)
$15\%$; uniaxial tensile strains along the armchair direction with the strain of (d) $5\%$, (e) $9\%$ and (f) $15\%$, and along the zigzag direction with the strain of
(g) $5\%$, (h) $15\%$ and (i) $19\%$.}
\end{figure}

It is known that graphdiyne is composed of acetylenic linkages (two linked \ce{{C}\tbond{C}} bonds) and carbon hexagons. The distributions of electron densities of graphdiyne under different tensile strains are given in Fig. 6. The distribution of electron density of graphdiyne without a strain is shown in Fig. 6(a). As one may see, the electron densities on carbon
hexagons are similar to those of graphene, which display a six-fold
symmetry. For acetylenic linkages, the electron density of
\ce{{C}\tbond{C}} bonds is obviously higher than that of the
C$\mbox{--}$C bonds. As shown in Figs. 6(b) and (c), with increasing the
symmetrical biaxial tensile strain, the electron densities on carbon
hexagons keep the symmetry, but the densities between carbon atoms
become smaller. In acetylenic linkages, the electron densities on
\ce{{C}\tbond{C}} bonds keep unchanged, while those of C$\mbox{--}$C bonds
decrease significantly, showing that C$\mbox{--}$C bonds become weak on carbon
hexagons and acetylenic linkages, which may cause the energy
bands more narrow and flat, leading to the band gaps become larger. The
electron densities of graphdiyne under the uniaxial tensile strain along the armchair direction are presented in Figs. 6(d)-(f). As the carbon linkages along the armchair direction are elongated, the
electron densities on C$\mbox{--}$C bonds decrease while those of
$\ce{{C}\tbond{C}}$ bonds change little. The carbon linkages along other
directions are compressed, where the electron densities on C$\mbox{--}$C
and $\ce{{C}\tbond{C}}$ bonds are still high and even slightly larger than those of the unstrained situation. Figs. 6(g) and (h) reveal the electron densities of
graphdiyne under the uniaxial tensile strain of $5\%$ and $15\%$ along the zigzag direction, respectively. In contrast to the case along the armchair direction, the carbon linkages along the armchair direction are compressed, and the electron densities are slightly larger than those of the unstrained situation, while the carbon linkages along the zigzag direction are elongated, leading to the electron
densities on C$\mbox{--}$C bonds decrease. However, when the strain along the zigzag direction is applied up to $19\%$ (Fig. 6(i)), the electron densities on carbon hexagons and C$\mbox{--}$C bonds decrease dramatically, and a large number
of electrons are localized on $\ce{{C}\tbond{C}}$ bonds, which
may be the reason that the energy gap of graphdiyne opens again.

\begin{figure}[tbp]
\includegraphics[width=6 cm]{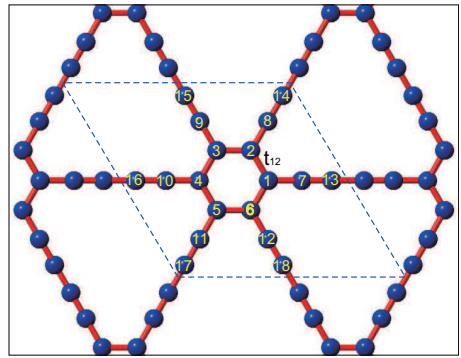}
\caption{(Color online) The atom sites $i$ (from 1 to 18) in one
unit cell (the parallelogram) of graphdiyne are labeled with numbers, and
$t_{i,j}$ in Eq. (1) is defined as the hopping amplitude between
sites $i$ and $j$.}
\end{figure}

To understand the DFT results of the electronic
properties of graphdiyne under different tensile strains, a tight-binding (TB) model of $\pi$ electrons in graphdiyne is proposed near the Fermi level. As indicated above in PDOS analyses, the electronic states around the Fermi level are predominantly contributed by $2p_{z}$ orbitals of carbons in graphdiyne, no matter whether strains are applied or not. Thus, only $\pi$ electrons and nearest
neighbor hoppings are sufficient to describe the low-lying properties of electrons in graphdiyne. The TB Hamiltonian could be written as
\begin{equation}
H = \sum_{\langle i,j \rangle} t_{i,j}
(\hat{c}^{\dagger}_{i} \hat{c}_{j} + h.c.)
\end{equation}
where the operators $\hat{c}^{\dagger}_{i}$ and ${c}_{i}$ are the
creation and annihilation operators of $\pi$ electrons at site $i$, ${\langle i,j
\rangle}$ denotes the sum over nearest neighbors within the unit cell, and $t_{i,j}$ stands for the hopping amplitude between sites $i$ and $j$.

As shown in Fig. 7, there are 18 atoms in one unit cell. As
$i$ and $j$ should go through all of these atoms, the matrix form of Hamiltonian is 18$\times$18, where the number of independent hopping
amplitude $t_{i,j}$ relies on the symmetry of the unit cell. We first
calculate the TB band structure of graphdiyne without a strain, as
given in Fig. 3(a). One may see that the TB results are in good agreement with the DFT calculations, implying that the low-energy properties of
$\pi$ electrons in grpahdiyne can be well described by the TB
Hamiltonian with four independent hopping amplitude $t_{i,j}$. When the biaxial tensile strain is applied to graphdiyne, the corresponding TB band structures are found very well consistent with the DFT results (Figs. 3(b)-3(d)), where there is also 4 independent $t_{i,j}$ parameters owing to the same symmetry as in the primitive graphdiyne. We also found that the fitting values of $t_{i,j}$ become smaller with increasing the strain, and $t_{i,j}$ on C$\mbox{--}$C bonds are much smaller than those on $\ce{{C}\tbond{C}}$ bonds, which are also consistent with the
decrease of electron densities shown in Figs. 6(b) and (c).

For the cases of graphdiyne under uniaxial tensile strains, there are eight independent parameters $t_{i,j}$. The corresponding TB band structures are shown in
Figs. 4 and 5, which give the results in well agreement with the DFT calculations. In these cases the degeneracies of the lowest conduction bands and the highest valence bands are lifted, and the band gaps become narrow. For the case of strains applied along the armchair direction, $t_{i,j}$ along this direction are smaller than those along the other directions, being consistent with the electron densities shown in Figs. 6(d)-(f). The minimal band gap is confirmed at $\Gamma$ point, and decreases to nearly zero. The case of tensile strains applied along the zigzag direction also recovers the DFT results (Figs. 6(g)-(i)). The TB calculations support as well that the broken geometrical symmetry of graphdiyne caused by the asymmetrical uniaxial strains lifts the degeneracy of energy bands, resulting in the shrink of the band gap.

By means of first-principles calculations combined with the TB approximations, the effect of symmetrical biaxial and asymmetrical uniaxial strains on
electronic properties of graphdiyne is explored, which is shown remarkably different from that on graphene. The calculated results reveal that the symmetrical biaxial
strain can enlarge the band gap of graphdiyne, while the
asymmetrical uniaxial strains could remove the band degeneracies, giving rise to the decrease of the band gap. It is found that the electronic structures around the Fermi level are dominantly determined by $2p_{z}$ orbitals of $\pi$ electrons. The strain-induced semiconductor-semimetal transition in graphdiyne is observed when the asymmetrical uniaxial tensile strain is applied, and Dirac cone-like electronic
structures are obtained. This study provides a possible way to modulate the
electronic properties of 2D materials like graphdiyne through external strains,
which may have potential applications.


We are grateful to Eric Germaneau, Z. C. Wang, S. Y. Yue and K. H. Zhang for useful discussions. All calculations were completed in Shanghai Supercomputer Center, China.
This work is supported in part by the NSFC (Grants No. 90922033, No. 10934008, No.
10974253 and No. 11004239), the MOST of China (Grant No.
2012CB932900, 2013CB933401), and the CAS.

\end{document}